\documentstyle[12pt]{article}

\newcommand{\be}{\begin{equation}}
\newcommand{\ee}{\end{equation}}
\newcommand{\binomial}[2]{\left(\begin{array}{c}#1\\#2\end{array}\right)}
\newcommand{\half}{{1\over 2}}

\advance\oddsidemargin by -\textwidth
\advance\oddsidemargin by -1in
\evensidemargin=\oddsidemargin

\begin{document}
$\mbox{ }$
\vspace{-3cm}
\begin{flushright}
\begin{tabular}{l}
{\bf KEK-TH-432 }\\
%{\bf KEK preprint ??? }\\
March 1995
\end{tabular}
\end{flushright}

\vspace{2cm}
\begin{center}
\Large
{\baselineskip26pt \bf On the entropy bound of three dimensional \\
				simplicial gravity}
\end{center}
\vspace{1cm}
\begin{center}
\large
{\sc Tsuguo Mogami}
\end{center}
\normalsize
\begin{center}
%\begin{tabular}{l}
{\it Department of Physics, Kyoto University,}\\
{\it Kitashirakawa, Kyoto 606, Japan}\\
		{\it and} \\
{\it KEK Theory Group, Tsukuba, Ibaraki 305, Japan}\\
%\end{tabular}
\end{center}
\vspace{2cm}
\begin{center}
\normalsize
ABSTRACT
\end{center}
{\rightskip=2pc %3pc
\leftskip=2pc %3pc
\normalsize
It is proven that the partition function of 3-dimensional simplicial
gravity has an exponential upper bound with the following assumption:
any three dimensional sphere $S^3$ is constructed
by repeated identification of neighboring links and neighboring triangles
in the boundary of a simplicial 3-ball.  This assumption is weaker than the
one proposed by other authors.
\vglue 0.6cm}

\newpage

The remarkable success of two dimensional quantum gravity owes itself to
the interplay of continuum theory and discretized theory called ``dynamical
triangulation''.  This success tempted people to formulate a discretized
theory of higher dimensional quantum gravity.  Among such theories,
simplicial quantum gravity\cite{ADJ} is the most natural generalization of
the dynamical triangulation or the matrix model.  In this short note we
restrictour attention to the three dimensional model.

In the model of simplicial gravity, the path integral is defined as the sum
over all the possible configurations of complexes consisting of equilateral
simplexes.  The discretized version of the Einstein-Hilbert
action for a triangulation $T$ is given by
\be
\int d^3x\ \sqrt g R \sim a\ \{2\pi N_1(T) - 6 \cos^{-1}({1\over3})
\times N_3(T) \},
\ee
where $a$ is the length of the links. $N_i(T)$ is the number of
$i$-simplexes contained in the triangulation $T$. The discretized
cosmological term is written as $\lambda N_3(T)$, where $\lambda$ denotes
the bare cosmological constant.  Using the action (eq(1))
and the relation
$N_0(T)-N_1(T)+N_2(T)-N_3(T)=0$, which is equivalent to saying $T$ is a
manifold, the partition function is written as
\be
Z = \sum_T {\rm e}^{\alpha N_0(T) -\kappa N_3(T)} \equiv \sum_{N_3}
Z(\alpha,N_3) {\rm e}^{-\kappa N_3},	\label{PartitionFn}
\ee
where $\alpha$ is a constant.  Here the topology of $T$ is fixed, for
instance, to be $S^3$.  The two dimensional analogue of this partion
function had remarkable success.

We expect $Z(\alpha,N_3)$ to grow as ${\rm e}^{\kappa_{\rm c}N_3}$ with its
volume $N_3$ as in the case of two dimensions.  Here $\kappa_{\rm c}$ is a
certain constant.  Then we may obtain the continuum limit by letting
$\kappa \to \kappa_{\rm c}$.  If $Z(\alpha,N_3)$ grows faster than
exponential, the partion function (\ref{PartitionFn}) is ill-defined.  We
must show this exponential growth analytically.  (Numerical simulations
give indication of exponential growth\cite{AV,CKR}.)

Though some attempts\cite{Bou,DJ,BBCM} have been made to prove that the
partition function has an exponential upper bound, they are not completed
yet.  To prove this, it is conjectured that any simplicial manifold with
the topology of $S^3$ is ``locally constructible'' in ref.\cite{Bou,DJ}.
If this conjecture is right, the exponential bound is proven.  The
terminology ``local constructibility'' of a manifold means that we can
construct the manifold by successively identifying neighboring pairs of
triangles on the boundary of a simplicial ball.

In this short note, we discuss whether the condition of local
constructibility may be loosened without spoiling an exponential bound.
Then the conjecture of local constructibility of $S^3$ is substituted by a
weaker one.  This weaker version of the conjecture might be easier to
prove, or it is possible that this weaker conjecture survives even if the
original conjecture fails.

We first examine the original version of local construction proposed in
ref.\cite{Bou,DJ}.We say a manifold $\cal M$ is locally constructible, if
there exist a sequence of simplicial manifolds $T_1,T_2,\cdots,T_n$ such
that
\begin{itemize}
\item[i)] $T_1$ is a tetrahedron.
\item[ii)] $T_{i+1}$ is constructed from $T_i$ by
	\begin{itemize}
	\item[a)] gluing a new tetrahedron to $T_i$ along one of the
triangles on the boundary of $T_i$, or
	\item[b)] identifying a pair of triangles in $\partial T_i$ which
share at least one link in $\partial T_i$.
	\end{itemize}
\item[iii)] $T_n={\cal M}$.
\end{itemize}

It is not difficult to show that a locally constructible closed 3-d
manifold is a three sphere \cite{DJ}.  Though the local constructibility of
all the $S^3$ manifold is not proven, proving this conjecture seems to be
the most promising way to prove an exponential bound.  Assuming the local
constructibility, an exponential bound is shown in the following way.

We reorder the construction so that all the tetrahedra are assembled before
doing the construction ii)b).  If a manifold contains $N$ tetrahedra, $T_N$
is a tree-like manifold with the topology of $B^3$.  The number of distinct
configurations of such manifolds is bounded by ${C_1}^N$ where $C_1$ is a
certain constant.

Then the series $T_N,T_{N+1},\cdots,T_{2N+2}$ is obtained by successively
applying ii)b) to $T_N$.  For each of configuration possible at $T_N$,
there may be
many ways to identify the triangles on the boundary.  We are going to
estimate the number of these ways from above.  Consider an imaginary
surface which has the same triangulations as $\partial T_N$.  This surface
consists of $2N+2$ triangles, and is used to remember how the
identification is done.  Suppose triangles $A$ and $B$ on $\partial T_N$
sharing a link $L$ are identified in the process $T_N \to T_{N+1}$.  In
parallel with this, we mark two arrows pointing $L$ on $A$ and $B$ in the
imaginary surface (fig1.a).  If the next identification $T_{N+1} \to
T_{N+2}$ is, for example, identifying triangles $C$ and $D$ which share a
link on the surface of $T_{N+1}$, we draw arrows pointing the link on the
imaginary surface (fig1.b).  (The shared link is two links on the imaginary
surface.)

Every successive identification $T_N \to T_{2N+2}$ may be interpreted as a
configuration of arrows.  Conversely, if we have a configuration of arrows,
we can specify how the identification is done.  Though there exist
configurations of arrows which can not be interpreted as successive
identifications of neighboring triangles, the number of ways to identify
all the triangles has an upper bound $3^{2N+2} \equiv 9{C_2}^N$, which is
the number of possible configurations of arrows.
(In ref.\cite{DJ} $C_2$ is estimated to be 384.)
Combining these bounds, the number of locally constructible triangulations
of $S^3$ is bounded by $C^N$ where $C=C_1 C_2$.

Here we present a weaker version of local constructibility by adding
another construction process:
\begin{itemize}
\item[ii)c)] Identify some links which share a common point.  No link
participates in this process twice.
\end{itemize}
This process is characterized by a point sitting at the center of the
links identified.  We call
such a point ``center point'' here.  It might be easier to prove this
looser
version of local constructibility than to prove the original one.

Let us prove that an exponential bound holds even if we add the
construction process ii)c).  There are many ways to construct the same
complex.  For example, before identifying two neighboring triangles we may
identify its sides.  Using such freedom, it is possible to arrange the
construction process so that we do all the ii)a)'s in the first $N$ steps
and all the ii)b)'s in the last $N+1$ steps.  The number of distinct
identifications in the last $N+1$ process is bounded again by $9{C_2}^N$.
Hence, an exponential bound is shown if the number of different ways to
identify links by repeated ii)c) is bounded exponentially.

Before counting the number of different configurations obtained by repeated
ii)c), we must know the number of configurations after a single ii)c).
Suppose a case where there are 6 links sharing a point $p$ (fig.2).  One of
these links is not identified (link $a$ in fig.2).  The link $c$ is not
identified now (because the arrow is not pointing $p$).  The remaining four
links are identified.  There are three ways to identify these links.  The
first possibility is that these four are identified altogether.  The second
is that link $b$ and $d$ are identified while $e$ and $f$ are identified
separately.  The third is pairing $b$ with $f$ and $d$ with $e$.  The other
ways of pairing must not be done because the complex will not become a
manifold.

Now we go to the generic case where $m$ links sharing a point are
identified.  Let us denote by $N(m)$ the number of different ways to
identify these links.  Every one of these link identification is
represented by a graph having $m$ legs.  An example of such a graph is
shown in fig.3.  These graphs must be planar because $\cal M$ must be a
manifold.  If a graph is rotated, it represents a different identification.
So, one of these legs must be marked.  $N(m)$ is obtained by counting the
number of such graphs having $m$ legs.  This may be calculated by an
equation which looks like the Schwinger-Dyson equation for the matrix
model.
\be
N(m) = \delta_{m,0}\ +\sum_{m_1+m_2+2=m} N(m_1) N(m_2+1)\ +
\sum_{m_1+m_2+2=m} N(m_1) N(m_2),
\ee
where we defined $N(0)$ to be 1 and $N(1)$ to be 0 and the indices
$m_1,m_2$ in the sum run over non-negative integers.  The second term in
the equation above is for the case in which the marked leg is connected
to two or more legs, and the third term is for the case in which marked leg
is connected to one other leg.
It is easily shown that
\be
N(m) \leq 4^m.
\ee
This estimation is rather loose, but enough to show an exponential bound in
the final expression.

Now we consider the whole link identification.  Draw arrows on each links
to show which is the ``center point'' for the link.  Suppose $M$ links are
identified in the repeated application of ii)c).  The number of different
ways to draw the arrows is $_{N_1}{\rm C}_M 2^M$.  From here on we define
$N_i$ to be the number of $i$-simplexes on $\partial T_N$.

We next count the number of possible ways to identify links for each of
configuration of arrows.  Some of the identification may be done
independently.
Identify all the ii)c)'s which may be identified independently as the first
round.  The number of different ways to take the points at the center of
identification is bounded by
\be
\binomial{N_0}{n_1},
\ee
where $n_1$ is the number of the center points.
Suppose $m_1$ links are identified in these $n_1$ identifications ii)c).
We denote by $m_{1,p}$ the number of links which have an arrow pointing $p$.
It holds that
\be
\sum_p m_{1,p} = m_1.
\ee
$p$ runs over the $n_1$ center points.
The number of configurations after identification of these $m_1$ links is
bounded by
\be
\prod_p N(m_{1,p}) \leq \prod_p 4^{m_{1,p}} = 4^{m_1}.
\ee

Then, we go to the next round of identification.  Identify all the ii)c)'s
which may be identified independently as we did in the first round.
Let us denote by $n_2$
the number of center points
in this round. $n_2$ is bounded by $[m_1/2]$, where $[\cdots]$ means the
largest integer smaller or equal to the number in it.  Let's suppose that
$m_2$
links are identified in this round.  The number of distinct ways of
identification in this round is bounded by
\be
\binomial{[\half m_1]}{n_2}\ \prod_p N(m_{2,p}) .
\ee
The index $p$ in the equation above runs all over the $n_2$ points.  We
have $\sum_p m_{2,p} = m_2$ also.

Repeating the same procedure, we finally obtain a bound for the link
identification.  The number of different ways of link identification is
bounded by
\def\nn{\nonumber \\}
\begin{eqnarray}
\lefteqn{
	\sum_{M=1}^{N_1} \binomial{N_1}{M} 2^M
	\sum_{f=1}^M \sum_{m_1+m_2+\cdots+m_f=M}
	\binomial{N_0}{n_1} 4^{m_1} \binomial{[\half m_1]}{n_2} 4^{m_2}
%	\binomial{[m_2/2]}{n_3} 4^{m_3}
	\cdots \binomial{[\half m_{f-1}]}{n_f} 4^{m_f}
	}	\nn
&\leq& \sum_{M=1}^{N_1} \binomial{N_1}{M} 2^M
	\sum_{f=1}^M \binomial{M-f-1}{f-1} 2^{N_0}
	2^{M/2} 4^{M}	\\
%	2^{(m_1+\cdots+m_{f-1})/2} 4^{m_1+\cdots+m_f}	\\
%&\leq& \sum_{M=1}^{N_1} \binomial{N_1}{M} 2^M
%	\sum_{f=1}^M \binomial{M-f-1}{f-1} 2^{N_0} 2^{{5\over2}M} \nn
&\leq& \sum_{M=1}^{N_1} \binomial{N_1}{M} 2^{{9\over2}M+N_0}
	\ = \ 2^{N_0} (2^{9\over2}+1)^{N_1}
	\ = \ 2^{2+{N_2\over2}} (2^{9\over2}+1)^{{3\over2}N_2}. \nonumber
\end{eqnarray}
Here we finally obtain $C_2 = 9\times2^{11N_1/2+N_0}$.  We now know that
the number of manifolds with a weaker version local constructibility is
bounded by $C^N$, where $C=C_1C_2$.

In this short note we have discussed an extension of local construction
which might serve as a way to show an exponential bound of 3-spheres.  It
might be possible that this version of local constructibility is easier to
prove, or may survive even if original local constructibility fails.

The author thanks N. Ishibashi and B. Bullock for reading the manuscript.

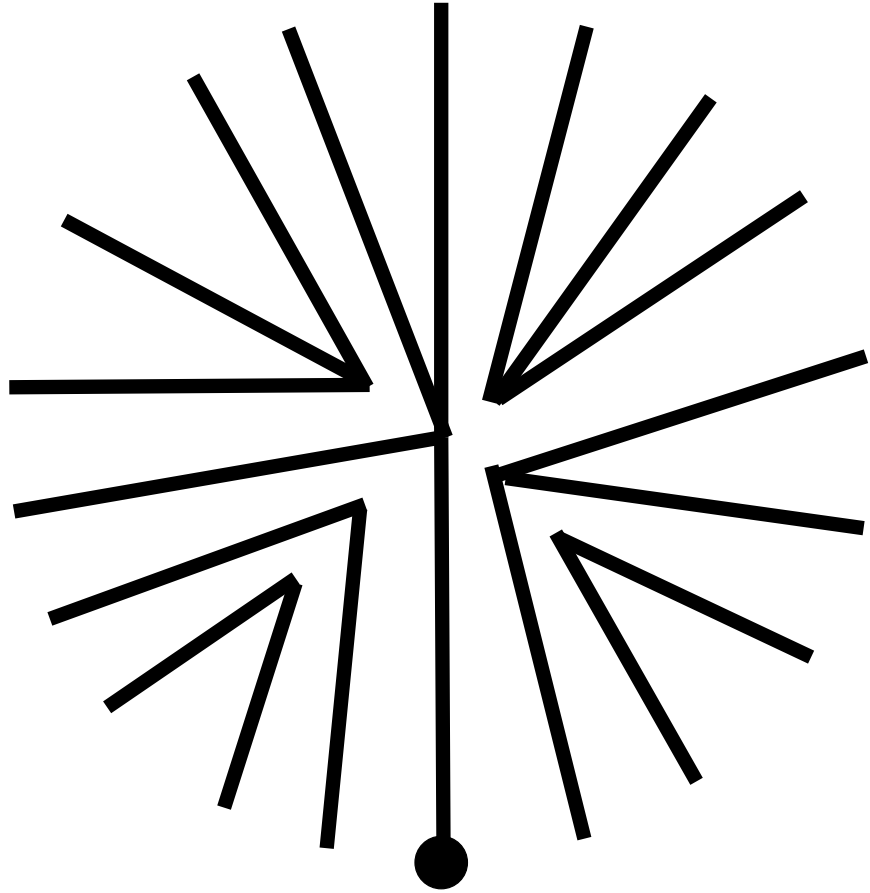
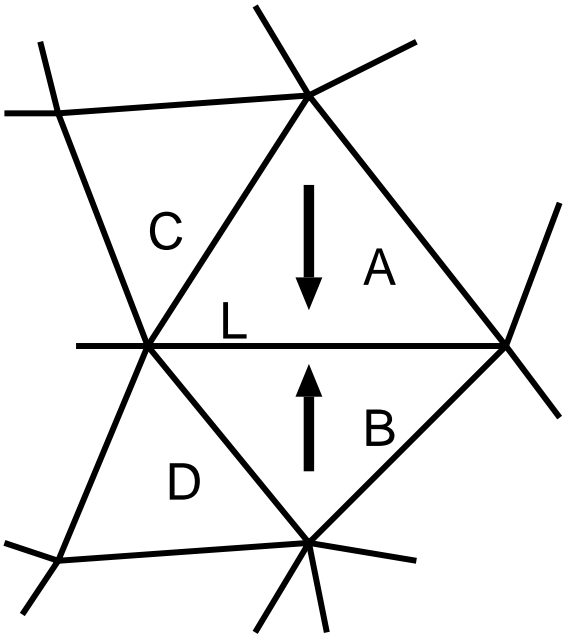
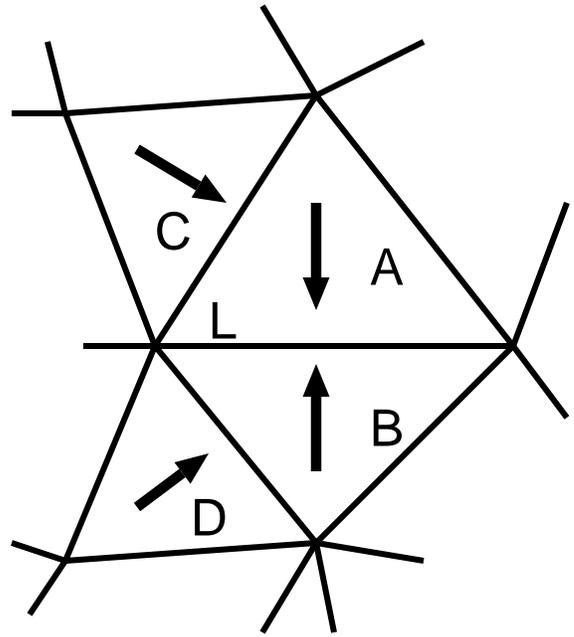
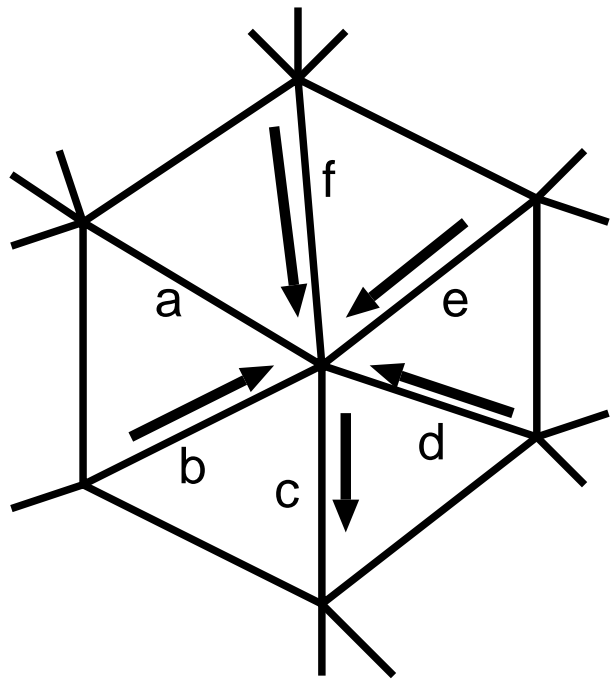
\end{document}